\documentstyle[prl,eqsecnum,aps]{revtex}

\begin{document}
\author{Jian-Qi Shen\footnote{E-mail address: jqshen@coer.zju.edu.cn}$,$ Hong-Yi Zhu and Hong Mao}
\address{State Key Laboratory of Modern Optical Instrumentation, Center for Optical\\
and Electromagnetic Research, College of Information Science and Engineering%
\\
Zhejiang Institute of Modern Physics and Department of Physics,\\
Zhejiang University, Hangzhou 310027, People$^{,}$s Republic of China}
\date{\today}
\title{An approach to exact solutions of the time-dependent supersymmetric
two-level three-photon Jaynes-Cummings model}
\maketitle

\begin{abstract}
By utilizing the property of the supersymmetric structure in the two-level
multiphoton Jaynes-Cummings model, an invariant is constructed in terms of
the supersymmetric generators by working in the sub-Hilbert-space
corresponding to a particular eigenvalue of the conserved supersymmetric
generators. We obtain the exact solutions of the time-dependent
Schr\"{o}dinger equation which describes the time-dependent supersymmetric
two-level three-photon Jaynes-Cummings model (TLTJCM) by using the
invariant-related unitary transformation formulation. The case under the
adiabatic approximation is also discussed.

\end{abstract}

\section{Introduction}

The interaction between a two-level atom and a quantized single-mode
electromagnetic field can be described by the Jaynes-Cummings model (JCM)
\cite{Jaynes} which has been applied to investigate many quantum effects
such as the quantum collapses and revivals of the atomic inversion, photon
antibunching, squeezing of the radiation field, inversionless light
amplification, electromagnetic induced transparency\cite
{Eberly,Alexanian,Wodkiewicz,Imamolglu}, etc.. In addition to the standard
JCM, there exists another type of JCM which possesses supersymmetric
structure\cite{Sukumar,Kien}. In this generalization of the J-C model, the
atomic transitions are mediated by $k$ photons. Singh has shown that this
model can be used to study multiple atom scattering of radiation and
multiphoton emission, absorption, and laser processes\cite{Singh}. Some
authors introduced a supersymmetric unitary transformation to diagonalize
the Hamiltonian of this supersymmetric JCM and obtain the eigenfunctions of
the stationary Schr\"{o}dinger equation$\cite{Lu1,Lu2}$. It is of great
interest to investigate the geometric phase factor of the time-dependent JCM
since the geometric phase factor\cite{Berry} appears only in systems with
the time-dependent Hamiltonian. It can be verified as shown in the present
paper that the exact solutions and the geometric phase factor of the
two-level JCM whose Hamiltonian has time-dependent parameters can also be
obtained by making use of the generalized invariant theory\cite{Gao1}. For
simplicity and convenience, in this paper we only investigate the
time-dependent supersymmetric two-level three-photon Jaynes-Cummings model
(TLTJCM). This method can also be generalized to the time-dependent JCM with
more than three photons.

The invariant theory which is appropriate for treating time-dependent
systems was first proposed by Lewis and Riesenfeld (L-R)\cite{Lewis} in
1969. In 1991, Gao et al. generalized the L-R invariant theory and proposed
the invariant-related unitary transformation formulation\cite{Gao1,Gao3}. In
this formulation the eigenstates of the time-dependent invariants are
replaced with those of the time-independent invariants through the unitary
transformation and the exact solutions (which contain the dynamical and
geometric phase factor) of the time-dependent Schr\"{o}dinger equation are
obtained. Many works have shown that the invariant-related unitary
transformation approach is a powerful tool for treating time-dependent
problems and geometric phase factor\cite{Fu1,Shen,Gxc}.

\section{The exact solutions of the time-dependent TLTJCM}

The Hamiltonian of the TLTJCM under the rotating wave approximation is given
by

\begin{equation}
H(t)=\omega (t)a^{\dagger }a+\frac{\omega _{0}(t)}{2}\sigma
_{z}+g(t)(a^{\dagger })^{3}\sigma _{-}+g^{\ast }(t)a^{3}\sigma _{+},
\label{eq31}
\end{equation}
where $a^{\dagger }$ and $a$ are the creation and annihilation operators for
the electromagnetic field, and obey the commutation relation $\left[
a,a^{\dagger }\right] =1$; $\sigma _{\pm }$ and $\sigma _{z}$ denote the
two-level atom operators which satisfy the commutation relation $\left[
\sigma _{z},\sigma _{\pm }\right] =\pm 2\sigma _{\pm }$ ; $g(t)$ and $%
g^{\ast }(t)$ are the coupling coefficients and $3$ is the photon number in
each atom transition process; $\omega _{0}(t)$ and $\omega (t)$ are
respectively the transition frequency and the mode frequency. All the
parameters in Eq. (\ref{eq31}) are time-dependent, and the time-dependent
Schr\"{o}dinger equation for this system is

\begin{equation}
i\frac{\partial \left| \Psi (t)\right\rangle _{s}}{\partial t}=H(t)\left|
\Psi (t)\right\rangle _{s}.  \label{eq24}
\end{equation}

The supersymmetric structure can be found in the TLTJCM by defining the
following supersymmetric transformation generators\cite{Lu1,Lu2}:

\begin{eqnarray}
N &=&a^{\dagger }a+\sigma _{z}+\frac{1}{2},\quad N^{^{\prime }}=\left(
\begin{array}{cc}
a^{3}(a^{\dagger })^{3} & 0 \\
0 & (a^{\dagger })^{3}a^{3}
\end{array}
\right) ,  \nonumber \\
Q &=&(a^{\dagger })^{3}\sigma _{-}=\left(
\begin{array}{cc}
0 & 0 \\
(a^{\dagger })^{3} & 0
\end{array}
\right) ,\quad Q^{\dagger }=a^{3}\sigma _{+}=\left(
\begin{array}{cc}
0 & a^{3} \\
0 & 0
\end{array}
\right) .  \label{eq32}
\end{eqnarray}
It is easily verified that $(N,N^{^{\prime }},Q,Q^{\dagger })$ form
supersymmetric generators and have supersymmetric Lie algebra properties,
i.e.,

\begin{eqnarray}
Q^{2} &=&(Q^{\dagger })^{2}=0,\quad \left[ Q^{\dagger },Q\right]
=N^{^{\prime }}\sigma _{z},\quad \left[ N,N^{^{\prime }}\right] =0,\quad %
\left[ N,Q\right] =Q,  \nonumber \\
\left[ N,Q^{\dagger }\right] &=&-Q^{\dagger },\quad \left\{ Q^{\dagger
},Q\right\} =N^{^{\prime }},\quad \left\{ Q,\sigma _{z}\right\} =\left\{
Q^{\dagger },\sigma _{z}\right\} =0,  \nonumber \\
\left[ Q,\sigma _{z}\right] &=&2Q,\quad \left[ Q^{\dagger },\sigma _{z}%
\right] =-2Q^{\dagger },\quad \left( Q^{\dagger }-Q\right) ^{2}=-N^{^{\prime
}},  \label{eq33}
\end{eqnarray}
where $\left\{ {}\right\} $ denotes the anticommuting bracket. By the aid of
Eqs. (\ref{eq32}) and (\ref{eq33}), the Hamiltonian (\ref{eq31}) of the
TLTJCM can be rewritten as

\begin{equation}
H(t)=\omega (t)N+\frac{\omega _{0}(t)-2\omega (t)}{2}\sigma
_{z}+g(t)Q+g^{\ast }(t)Q^{\dagger }-\frac{\omega (t)}{2}.  \label{eq34}
\end{equation}

According to the L-R invariant theory\cite{Lewis}, one should first
construct an invariant $I(t)$ in order to show the solvability of Eq. (\ref
{eq24}). A Hermitian operator $I(t)$ is called invariant if it satisfies the
following invariant equation

\begin{equation}
\frac{\partial I(t)}{\partial t}+\frac{1}{i}[I(t),H(t)]=0,  \label{eq21}
\end{equation}
and the eigenvalue equation of the time-dependent invariant is given by
\begin{equation}
I(t)\left| \lambda _{n},t\right\rangle =\lambda _{n}\left| \lambda
_{n},t\right\rangle ,  \label{eq221}
\end{equation}
where $\frac{\partial \lambda _{n}}{\partial t}=0.$ It can be seen from the
invariant equation (\ref{eq21}) that $I(t)$ is the linear combination of $%
N,\sigma _{z},Q$ and $Q^{\dagger }.$ However, it should be emphasized that
the generalized invariant theory\cite{Gao1} can only be applied to study the
system with the quasialgebra defined in\cite{Mizrahi}. Unfortunately, there
is no such quasialgebra for the TLTJCM in Eq. (\ref{eq33}). Many problems
have been solved in Quantum Mechanics by working in the sub-Hilbert-space
corresponding to a particular eigenvalue of the Hamiltonian \cite{Schiff}.
We show that in the case of the TLTJCM, a generalized quasialgebra, which
enables one to obtain the complete set of the exact solutions for the
TLTJCM, can also be found by working in a sub-Hilbert-space corresponding to
a particular eigenvalue of the supersymmetric generator $N^{^{\prime }}$.

Using $a^{3}(a^{\dagger })^{3}\left| m\right\rangle =\frac{(m+3)!}{m!}\left|
m\right\rangle $ and $(a^{\dagger })^{3}a^{3}\left| m+3\right\rangle =\frac{%
(m+3)!}{m!}\left| m+3\right\rangle ,$one can arrive at

\begin{equation}
N^{^{\prime }}%
{\left| m\right\rangle  \choose \left| m+3\right\rangle }%
=\lambda _{m}%
{\left| m\right\rangle  \choose \left| m+3\right\rangle }%
\label{eq35}
\end{equation}
with $\lambda _{m}=\frac{(m+3)!}{m!}.$ One thus obtains the supersymmetric
quasialgebra $(N,Q,Q^{\dagger },\sigma _{z})$ in the sub-Hilbert-space
corresponding to the particular eigenvalue $\lambda _{m}$ of $N^{^{\prime
}}, $ by replacing the generator $N^{^{\prime }}$ with $\lambda _{m}$ in the
commutation relations of Eq. (\ref{eq33}), namely,

\begin{equation}
\left[ Q^{\dagger },Q\right] =\lambda _{m}\sigma _{z},\quad \left\{
Q^{\dagger },Q\right\} =\lambda _{m},\quad \left( Q^{\dagger }-Q\right)
^{2}=-\lambda _{m}.  \label{eq36}
\end{equation}

In accordance with the invariant theory, we chose the invariant $I(t)$ to be
of the form

\begin{equation}
I(t)=-\frac{\sin \theta }{\lambda _{m}^{\frac{1}{2}}}[\exp (-i\phi )Q+\exp
(i\phi )Q^{\dagger }]+\cos \theta \sigma _{z},  \label{eq37}
\end{equation}
where $\theta $ and $\phi $ are time-dependent parameters. Substitution of
the expressions (\ref{eq34}) and (\ref{eq37}) for $I(t)$ and $H(t)$ into Eq.
(\ref{eq21}) leads to the following set of auxiliary equations
\begin{eqnarray}
\dot{\theta}\cos \theta \exp (-i\phi )-i\dot{\phi}\sin \theta \exp (-i\phi
)+i[(3\omega -\omega _{0})\sin \theta \exp (-i\phi )-2g\lambda _{m}^{\frac{1%
}{2}}\cos \theta ] &=&0,  \nonumber \\
\dot{\theta}-i\lambda _{m}^{\frac{1}{2}}[g\exp (i\phi )-g^{\ast }\exp
(-i\phi )] &=&0,  \label{eqq1}
\end{eqnarray}
where the dot denotes the time derivative. The two time-parameters $\theta $
and $\phi $ in $I(t)$ are determined by these two auxiliary equations.

Using the invariant-related unitary transformation method\cite{Gao1}, we
define the unitary transformation operator as follows

\begin{equation}
V(t)=\exp [\beta (t)Q-\beta ^{\ast }(t)Q^{\dagger }]  \label{eq39}
\end{equation}
with $\beta ^{\ast }(t)$ being the complex conjugation of $\beta (t).$ With
the help of the commutation relations (\ref{eq33}) and by the complicated
and lengthy computations, it can be found that if $\beta (t)$ and $\beta
^{\ast }(t)$ satisfy the following equations

\begin{equation}
\beta =-\frac{\frac{\theta }{2}\exp (-i\phi )}{\lambda _{m}^{\frac{1}{2}}}%
,\quad \beta ^{\ast }=-\frac{\frac{\theta }{2}\exp (i\phi )}{\lambda _{m}^{%
\frac{1}{2}}},  \label{eq310}
\end{equation}
then the following time-independent invariant $I_{V}$ can be obtained

\begin{equation}
I_{V}\equiv V^{\dagger }(t)I(t)V(t)=\sigma _{z}.  \label{eq311}
\end{equation}
Correspondingly, the Hamiltonian (\ref{eq34}) can be transformed into

\begin{eqnarray}
H_{V}(t) &\equiv &V^{\dagger }(t)H(t)V(t)-V^{\dagger }(t)i\frac{\partial }{%
\partial t}V(t)  \nonumber \\
&=&\omega N+\frac{\omega }{2}(\sigma _{z}-1)+\{-\frac{1}{2}\lambda _{m}^{%
\frac{1}{2}}[g\exp (i\phi )+g^{\ast }\exp (-i\phi )]\sin \theta +  \nonumber
\\
&&+\frac{\omega _{0}-3\omega }{2}\cos \theta -\frac{\stackrel{.}{\phi }}{2}%
(1-\cos \theta )\}\sigma _{z}.  \label{eq314}
\end{eqnarray}
by using the Baker-Campbell-Hausdorff formula\cite{Wei}

\begin{equation}
V^{\dagger }(t)\frac{\partial }{\partial t}V(t)=\frac{\partial }{\partial t}%
L+\frac{1}{2!}[\frac{\partial }{\partial t}L,L]+\frac{1}{3!}[[\frac{\partial
}{\partial t}L,L],L]+\frac{1}{4!}[[[\frac{\partial }{\partial t}%
L,L],L],L]+\cdots
\end{equation}
with $V(t)=\exp [L(t)].$ Under this unitary transformation\ref{eq39} the
time-dependent Schr\"{o}dinger equation (\ref{eq24}) is then transformed
into the following form

\begin{equation}
i\frac{\partial \left| \lambda _{n},t\right\rangle _{s0}}{\partial t}%
=H_{V}(t)\left| \lambda _{n},t\right\rangle _{s0}  \label{eq210}
\end{equation}
where

\begin{equation}
\left| \Psi (t)\right\rangle _{s}=V(t)\left| \lambda _{n},t\right\rangle
_{s0}.  \label{eq2111}
\end{equation}
One may show that the particular solution $\left| \lambda
_{n},t\right\rangle _{s}$ of Eq.(\ref{eq24}) differs from the eigenfunction $%
\left| \lambda _{n},t\right\rangle $ of the invariant $I(t)$ only by a phase
factor $\exp [i\phi _{n}(t)]$. Then the general solution of the
Schr\"{o}dinger equation (\ref{eq24}) can be written as

\begin{equation}
\left| \Psi (t)\right\rangle _{s}=%
\mathop{\textstyle\sum}%
_{n}C_{n}\exp [i\phi _{n}(t)]\left| \lambda _{n},t\right\rangle ,
\label{eq25}
\end{equation}
where

\[
\phi _{n}(t)=\int_{0}^{t}\left\langle \lambda _{n},t^{^{\prime }}\right| i%
\frac{\partial }{\partial t^{^{\prime }}}-H(t^{^{\prime }})\left| \lambda
_{n},t^{^{\prime }}\right\rangle dt^{^{\prime }},
\]

\begin{equation}
C_{n}=\langle \lambda _{n},t=0\left| \Psi (t=0)\right\rangle _{s}.
\label{eq26}
\end{equation}
It is easy to verify that the particular solution $\left| \lambda
_{n},t\right\rangle _{s0}$ of the time-dependent Schr\"{o}dinger equation (%
\ref{eq210}) is different from the eigenfunction $\left| \lambda
_{n}\right\rangle $ of $I_{V}$ only by the same phase factor $\exp [i\phi
_{n}(t)]$ as that in Eq.(\ref{eq25}), i.e.,

\begin{equation}
\left| \lambda _{n},t\right\rangle _{s0}=\exp [i\phi _{n}(t)]\left| \lambda
_{n}\right\rangle .  \label{eq211}
\end{equation}
Substitution of $\left| \lambda _{n},t\right\rangle _{s0}$ of Eq. (\ref
{eq210}) into Eq. (\ref{eq211}) yields

\begin{equation}
-\dot{\phi}(t)\left| \lambda _{n}\right\rangle =H_{V}(t)\left| \lambda
_{n}\right\rangle ,  \label{eq212}
\end{equation}
which means that $H_{V}(t)$ differs from $I_{V}(t)$ only by a time-dependent
multiplying c-number factor. Then the particular solution of Eq. (\ref{eq210}%
) can be easily obtained by calculating the phase from Eq. (\ref{eq212}).

The eigenstates of $\sigma _{z}$ corresponding to the eigenvalue $\sigma =+1$
and $\sigma =-1$ are $%
{1 \choose 0}%
$ and $%
{0 \choose 1}%
,$ and the eigenstate of $N^{^{\prime }}$ is $%
{\left| m\right\rangle  \choose \left| m+3\right\rangle }%
$ corresponding to Eq. (\ref{eq35}). From Eq. (\ref{eq26}), (\ref{eq211}), (%
\ref{eq212}), we obtain two particular solutions of the time-dependent
Schr\"{o}dinger equation of the TLTJCM which can be written in the following
forms

\begin{equation}
\left| \Psi _{m,\sigma =+1}(t)\right\rangle _{s}=\exp \{\frac{1}{i}%
\int_{0}^{t}[\stackrel{.}{\varphi }_{d,\sigma =+1}(t^{^{\prime }})+\stackrel{%
.}{\varphi }_{g,\sigma =+1}(t^{^{\prime }})]dt^{^{\prime }}\}V(t)%
{\left| m\right\rangle  \choose 0}%
\label{eq315}
\end{equation}
where

\begin{eqnarray}
\stackrel{.}{\varphi }_{d,\sigma =+1}(t^{^{\prime }}) &=&(m+\frac{3}{2}%
)\omega (t^{^{\prime }})-\frac{1}{2}\lambda _{m}^{\frac{1}{2}%
}\{g(t^{^{\prime }})\exp [i\phi (t^{^{\prime }})]+g^{\ast }(t^{^{\prime
}})\exp [-i\phi (t^{^{\prime }})]\}\sin \theta (t^{^{\prime }})  \nonumber \\
&&+\frac{\omega _{0}(t^{^{\prime }})-3\omega (t^{^{\prime }})}{2}\cos \theta
(t^{^{\prime }})  \label{eq111}
\end{eqnarray}
and

\begin{equation}
\stackrel{.}{\varphi }_{g,\sigma =+1}(t^{^{\prime }})=-\frac{\stackrel{.}{%
\phi }(t^{^{\prime }})}{2}[1-\cos \theta (t^{^{\prime }})];  \label{eq112}
\end{equation}
and

\begin{equation}
\left| \Psi _{m,\sigma =-1}(t)\right\rangle _{s}=\exp \{\frac{1}{i}%
\int_{0}^{t}[\stackrel{.}{\varphi }_{d,\sigma =-1}(t^{^{\prime }})+\stackrel{%
.}{\varphi }_{g,\sigma =-1}(t^{^{\prime }})]dt^{^{\prime }}\}V(t)%
{0 \choose \left| m+3\right\rangle }%
\label{eq316}
\end{equation}
where

\begin{eqnarray}
\stackrel{.}{\varphi }_{d,\sigma =-1}(t^{^{\prime }}) &=&(m+\frac{3}{2}%
)\omega (t^{^{\prime }})+\frac{1}{2}\lambda _{m}^{\frac{1}{2}%
}\{g(t^{^{\prime }})\exp [i\phi (t^{^{\prime }})]+g^{\ast }(t^{^{\prime
}})\exp [-i\phi (t^{^{\prime }})]\}\sin \theta (t^{^{\prime }})  \nonumber \\
&&-\frac{\omega _{0}(t^{^{\prime }})-3\omega (t^{^{\prime }})}{2}\cos \theta
(t^{^{\prime }})  \label{eq113}
\end{eqnarray}
and

\begin{equation}
\stackrel{.}{\varphi }_{g,\sigma =-1}(t^{^{\prime }})=\frac{\stackrel{.}{%
\phi }(t^{^{\prime }})}{2}[1-\cos \theta (t^{^{\prime }})].  \label{eq114}
\end{equation}
Generally speaking, in Quantum Mechanics, solution with
chronological-product operator (time-order operator) $P$, namely, where the
time-evolution operator $U(t)=P\exp [\frac{1}{i}\int_{0}^{t}H(t^{^{\prime
}})dt^{^{\prime }}],$is often called the formal solution. In the present
paper, however, the solution of the Schr\"{o}dinger equation governing a
time-dependent system is sometimes termed the explicit solution, for reasons
of the fact that it does not involve time-order operator. But, on the other
hand, by using Lewis-Riesenfeld invariant theory, there always exist
time-dependent parameters, for instance, $\theta $ and $\phi $ in this paper
which are determined by the auxiliary equations (\ref{eqq1}). Traditionally,
when employed in experimental analysis and compared with experimental
results, these nonlinear auxiliary equations should be solved often by means
of numerical computation. In view of the above reasons, the concept of
explicit solution is understood in a somewhat relative sense, namely, it can
be considered explicit solution when compared with the time-evolution
operator involving time-order operator; whereas, it cannot be considered
completely explicit solution for it is expressed in terms of some
time-dependent parameters which should be obtained via the auxiliary
equations. Hence, conservatively speaking, we regard the solution of the
time-dependent system presented in the paper as exact solution rather than
explicit solution.

These above two particular solutions of the Schr\"{o}dinger equation (\ref
{eq24}) contain the corresponding dynamical phase factor $\exp [\frac{1}{i}%
\int_{0}^{t}\stackrel{.}{\varphi }_{d,\sigma }(t^{^{\prime }})dt^{^{\prime
}}]$ and geometric phase factor $\exp [\frac{1}{i}\int_{0}^{t}\stackrel{.}{%
\varphi }_{g,\sigma }(t^{^{\prime }})]dt^{^{\prime }}$ with $\sigma =\pm 1.$
Apparently, it can be seen that the former phase factor depends on the
transition frequency $\omega _{0}(t)$ and the mode frequency $\omega (t)$ as
well as the coupling coefficients $g(t)$ and $g^{\ast }(t)$, whereas the
latter is immediately independent of these frequencies and the coupling
coefficients.

One of the theoretical applications of the exact solution of time-dependent
Schr\"{o}dinger equation is constructing the time-dependent coherent state
\cite{Xu}. For instance, one result in this paper may be given as follows
\begin{eqnarray}
\left| \Phi _{\sigma =+1}(t)\right\rangle &=&\exp (-\frac{\xi ^{2}}{2}%
)\sum_{m=0}^{\infty }\frac{\xi ^{m}}{\sqrt{m!}}\left| \Psi _{m,\sigma
=+1}(t)\right\rangle _{s}  \nonumber \\
&=&\exp (-\frac{\xi ^{2}}{2})\sum_{m=0}^{\infty }\frac{\xi ^{m}}{\sqrt{m!}}%
\exp \{\frac{1}{i}\int_{0}^{t}[\stackrel{.}{\varphi }_{\sigma
=+1}(t^{^{\prime }})]dt^{^{\prime }}\}V(t)%
{\left| m\right\rangle  \choose 0}%
\end{eqnarray}
with $\xi $ being a time-independent parameter and $\stackrel{.}{\varphi }%
_{\sigma =+1}(t)=\stackrel{.}{\varphi }_{d,\sigma =+1}(t)+\stackrel{.}{%
\varphi }_{g,\sigma =+1}(t).$ Time-dependent coherent state is believed to
be useful in investigating the classical properties of supersymmetric
Jaynes-Cummings model in quantum optics.

\section{Discussion under the adiabatic approximation}

The geometric phase factor in the adiabatic evolution of quantum systems was
first discussed by Berry\cite{Berry2}. Simon then showed that this phase is
in connection with a holonomy of the connection in the Hermitian line bundle
over the parameter space\cite{Simon}. This adiabatic phase, referred to as
the Berry phase, has attracted many attentions of the investigators in
various branches of physics\cite{Shapere}. Here we investigate the cases of
the adiabatic limit. Under the adiabatic limit, we assume that the time
derivative of $\theta $ vanishes, namely,

\begin{equation}
\dot{\theta}=0,  \label{eq120}
\end{equation}
then the following equations can be derived from the auxiliary equations (%
\ref{eqq1})

\begin{equation}
g=\left| g\right| \exp (-i\phi ),\quad g^{\ast }=\left| g\right| \exp (i\phi
),\quad (3\omega -\omega _{0}-\dot{\phi})\sin \theta =2\left| g\right|
\lambda _{m}^{\frac{1}{2}}\cos \theta .  \label{eq115}
\end{equation}
Inserting Eqs. (\ref{eq115}) into Eq. (\ref{eq37}), we obtain

\begin{equation}
I(t)=\frac{-2\cos \theta }{3\omega -\omega _{0}-\dot{\phi}}[g(t)Q+g^{\ast
}(t)Q^{\dagger }+\frac{-1}{2}(3\omega -\omega _{0}-\dot{\phi})\sigma _{z}].
\label{eq116}
\end{equation}
In accordance with the definition of an invariant, i.e., Eq. (\ref{eq21}),
one can draw a conclusion that an invariant can be regarded as the
Hamiltonian of the adiabatic-evolution system. Making a comparison between
Eq. (\ref{eq34}) and Eq. (\ref{eq116}), one can see that if
\begin{equation}
\dot{\phi}=\omega ,  \label{eq117}
\end{equation}
the following relation between $H(t)$ and $I(t)$ can be obtained

\begin{equation}
H(t)=\omega (t)N-\frac{\omega (t)}{2}-\frac{2\omega -\omega _{0}}{2\cos
\theta }I(t).  \label{eq118}
\end{equation}
Further analysis shows that for the general three-generator Lie-algebraic
systems, the invariant $I(t)$ is just considered the Hamiltonian $H(t)$
which can be expressed as
\begin{equation}
H(t)\sim I(t)  \label{eq119}
\end{equation}
in the adiabatic-evolution process. According to Eq. (\ref{eq21}) and Eq. (%
\ref{eq221}), the generator $N^{^{\prime }}$ is a time-independent
invariant, while $I(t)$ is a time-dependent invariant. Since the eigenstate
of $N^{^{\prime }}$ can be rewritten as

\begin{equation}
{\left| m\right\rangle  \choose \left| m+3\right\rangle }%
=\left| m\right\rangle
{1 \choose 0}%
+\left| m+3\right\rangle
{0 \choose 1}%
\end{equation}
which is apparently not the eigenstate of $I(t)$, even in the adiabatic
evolution, the problem of eigenvalue is not very necessary as in the
stationary Schr\"{o}dinger equation. An invariant which satisfies Eq. (\ref
{eq21}) is just a conserved operator. The product of two invariants, e.g., $%
I(t)N^{^{\prime }}$ is also an invariant\cite{Gao1}. More invariants can be
constructed in terms of

\begin{equation}
O(t)=U^{\dagger }(t)OU(t)
\end{equation}
where $O$ is an ordinary operator and $U(t)$ is the time-evolution operator
which is given

\begin{equation}
U(t)=V(t)\exp \{\frac{1}{i}\int_{0}^{t}[\stackrel{.}{\varphi }_{d,\sigma
}(t^{^{\prime }})+\stackrel{.}{\varphi }_{g,\sigma }(t^{^{\prime
}})]dt^{^{\prime }}\}.
\end{equation}
Here the invariant-related unitary transformation formulation provides the
evolution operator with an explicit expression rather than a formal solution
of equation

\begin{equation}
i\frac{\partial U}{\partial t}=H(t)U.
\end{equation}

It should be noted that, using Eq. (\ref{eq117}) and (\ref{eq120}), the
geometric phases $\int_{0}^{t}\stackrel{.}{\varphi }_{g,\sigma }(t^{^{\prime
}})]dt^{^{\prime }}$ can be rewritten as

\begin{equation}
\varphi _{g,\sigma }(t)=-\frac{\sigma }{2}(1-\cos \theta )\int_{0}^{t}\omega
dt  \label{eq121}
\end{equation}
with $\sigma =\pm 1.$ The geometric phases (or Berry phase) in a cycle
(i.e., one round trip) is

\begin{equation}
\varphi _{g,\sigma }(T)=-\frac{\sigma }{2}2\pi (1-\cos \theta ),
\label{eq122}
\end{equation}
where $2\pi (1-\cos \theta )$ is the solid angle over the parameter space of
the invariant $I(t),$ which unfolds the geometric meanings of the phase
factor. The right-hand side of Eq. (\ref{eq122}) is analogous to the
magnetic flux produced by a monopole of strength $-\frac{\sigma }{2}$
existing at the origin of the parameter space. This, therefore, implies that
geometric phase differs from dynamical phase and it involves the global and
topological properties of the time evolution of a quantum system.

\section{Concluding remarks}

In the present paper we have constructed an invariant in the
sub-Hilbert-space corresponding to a particular eigenvalue of the conserved
operator (time-independent invariant) $N^{^{\prime }}$ and obtained the
exact solutions of the time-dependent supersymmetric TLTJCM by making use of
the invariant-related unitary transformation formulation. This formulation
replaces eigenstates of the time-dependent invariants by those of the
time-independent invariants through the unitary transformation. In view of
the above calculation, we can see that this unitary transformation
formulation has some useful applications.

The exact solutions as well as their geometric phase factors of the
time-dependent single- and two-photon cases can be obtained by using the
present method. Since the three-level two-mode Jaynes-Cummings model plays
an important role in Quantum Optics\cite{Wu}, the supersymmetric structure
and the exact solutions of the time-dependent three-level two-mode
multiphoton JCM deserves further investigations by the formalism suggested
in the present paper. It is also interesting and necessary to obtain the
exact solutions of the supersymmetric TLTJCM without the rotating wave
approximation, by using this invariant-related unitary transformation
formulation.

Acknowledgments This project was supported by the National Natural Science
Foundation of China under the project No.$19775040$. The authors thank S.L
He for useful discussions and X.C.Gao for helpful suggestions.

\end{document}